# A Composite Target of a Radium Salt and a Soft Metal Matrix for Production of Ac-225 with a Proton or Electron Accelerator

W. T. Diamond (1), C. K. Ross (2) and H. A. Moore (3)

December 12, 2024

(1) Retired, Atomic Energy of Canada limited, Chalk River Laboratories, Chalk River, Ontario, Canada K0J 1J0, diamond_w45@yahoo.ca

(2) Retired, National Research Council, Ottawa, Ontario, Canada K1A 0R6, carlkross@gmail.com

(3) Retired, DuPont, Wilmington, Delaware, hamoore@telus.net

## Abstract

The production of $^{225}$Ac using either a proton or electron accelerator requires a target of $^{226}$Ra. Radium metal is difficult to work with and so a radium salt, such as radium carbonate, is preferred as a target material. Normally available as a powder, the average density of the powder is low and the thermal conductivity is poor, thus limiting the beam power that can be dissipated in the target. This work proposes the creation of a solid mixture of a radium salt powder with a metal matrix. Although aluminum powder has been used in similar applications, we suggest that indium powder is good choice for the metal matrix in the case of radioactive radium salts. The target can be formed with low die pressure without any need for heating, thus simplifying the hot-cell equipment needed for target preparation. We describe how the solid mixture can be formed, measure its thermal conductivity and compare the value to model estimates. We calculate the yield of $^{225}$Ac under different scenarios. Calculations show that the radioactive isotopes of indium produced during the irradiation should not produce significant handling challenges post-irradiation. For the proposed target geometry and beam parameters, thermal modelling indicates that the target temperature will be below the melting point of indium and the heat flux from the surfaces will be manageable. Thermal resistance at the target interfaces is shown to have a large effect on the target temperature. Using indium powder to form the mixture, occupying about 40 % of the target volume, we find that the yield of $^{225}$Ac is 80 GBq (2 Ci) for a 10-day irradiation with a 24 MeV, 3 kW proton beam and approximately the same for three milkings of $^{225}$Ra after a 10-day irradiation with a 25 MeV, 20 kW electron beam.

## 1. Introduction

There has been growing clinical evidence of the value of targeted alpha therapy for the treatment of several cancers[1]. The work has been slowed by the lack of availability of the key alpha emitting isotopes, especially $^{225}$Ac. There is an ongoing effort to produce new sources of $^{225}$Ac from several different accelerator-based routes that use radium as a target[2]. It can be produced with small cyclotrons with a proton energy of at least 16 MeV using the reaction $^{226}$Ra(p,2n)$^{225}$Ac and it can also be produced by the photonuclear reaction, $^{226}$Ra($\gamma$,n)$^{225}$Ra. The $^{225}$Ra decays via beta decay to $^{225}$Ac with a half-life of 14.9 d.

Radium is challenging to use as a target material and supplies are limited. It is a radioactive material with a specific activity of 37 GBq/g (1 Ci/g) and a high yield of penetrating gamma rays[3], and all handling activities must be done within a hot cell, even for target preparation. The first daughter product is $^{222}$Rn with a half-life of 3.8 d. This is a mobile noble gas that decays through multiple isotopes to $^{210}$Pb with a half-life of 22.2 y that decays via another alpha particle, posing alpha particle contamination challenges. Radium is a group 2 metal that is highly reactive with air or water and this makes electrodeposition of thin metal films in a hot cell a challenging approach to target production.

An alternative approach is to first process the radium to remove the daughter products that decay via gamma photons ($^{214}$Pb and $^{214}$Bi). The radium can then be handled in a shielded glove box. However, the daughter products will build in quickly, governed by the 3.82 d half-life of $^{222}$Rn. Processing will need to be done quickly, preferably within the same day.

Cyclotron targets will typically use a thin film of target material plated on an aluminum or copper backing and mounted in a target chamber at $10^{o}$ or less with respect to the incident proton beam to enable higher beam powers[4]. Once a metal radium layer is electrodeposited on a backing it must be kept under an inert gas and then be coated with a protective metallic film such as aluminum before it can be used as a target. The metallic coating needs to be thin because it will also intercept the proton beam at an angle of $10^{o}$ or less and present a thickness to the proton beam of nearly 10 times the actual thickness. However, it must also contain the large amount of radon isotopes produced by decay of the target material and co-produced by decay of isotopes produced during irradiation. It will also be subjected to internal pressure from the buildup of helium from the multiple alpha decays.

Targets for irradiation by electron bremsstrahlung need to be thicker because the bremsstrahlung is highly penetrating and produces fewer interactions per unit length than protons. This means the yield from a thin electrodeposited target will be low. Radium salts are more stable and can, in principle, be used as a thick target material suitable for irradiation with an electron accelerator. However, the thermal conductivity of powdered radium salts tends to be low (unknown, but likely in the range of 0.1 to 0.5 W/m-K as typical of powdered material[5]). This limits the power that can be deposited on the target material and the potential yield of $^{225}$Ac would be low.

A common method to improve the thermal conductivity of target materials with low thermal conductivities is to mix the target material with a metal matrix with high thermal conductivity and press the mixture with a high load to produce pellets or plates with an intermediate thermal conductivity[6].

Stolarz et al[7] prepared cyclotron targets with $CaCO_3$ powder mixed with graphite or aluminum powders. The mixtures are described as having equal amounts of the powders, which may mean by weight. The mixtures were compressed with a load of 40 kN on a small area. Heat transfer measurements were presented that showed improved heat transfer through both mixtures compared to compressed $CaCO_3$ with no filler, but no attempt was made to obtain a measurement of the thermal conductivity from the data. There were challenges noted in the separation of the compressed graphite/$CaCO_3$ target material because it was hard, limiting access of the acid used for target dissolving. The targets were irradiated with either protons or alphas at low currents of a few microamperes so the work does not provide insight to use at the high beam currents (typically 100 μA of protons or greater) that will be required to produce high yields of $^{225}$Ac.

This paper expands on the idea of using a metal matrix. Low molecular weight radium salts (to obtain the highest fraction of radium relative to the salt molecular weight) such as radium carbonate ($RaCO_3$) radium nitrate ($NO_3$), radium chloride ($RaCl_2$) or radium oxide ($RaO$), if it is shown to exist, could be used as a powder, and mixed with a soft metal powder to produce a composite target of reasonable thermal conductivity. Miniature mixer/shakers can be used to obtain good blending of dry powders of uneven sizes[8]. The mixture would then be compacted into a small puck that can be used as a target. The puck would be either produced in or installed into an aluminum can and a thin aluminum lid used to encapsulate the target. The composite target would have good thermal conductivity enabling reasonable power density during irradiation with either protons or electrons. The encapsulation enables the use of water cooling on both sides of the target to remove the heat.

Although aluminum powder has been used in similar applications, we propose to use indium metal powder, mixed with the radium salt with volume ratios in the compressed target of approximately 50 %, denoted by 50/50. Solid indium targets can be produced at low die pressures, thus simplifying the hot cell equipment needed. Tests with indium powder and $CaCO_3$ powder as a substitute for the radium salt (see the Appendix) have been conducted that show it to be a suitable choice. Tests were done at volume ratios of 33/67, 40/60 and 50/50 of the metal powder to $CaCO_3$ powder and the 40/60 ratio was chosen for more detailed tests that are reported in the Appendix.

## 2. Composite Target Technology

Composite targets are used for a variety of mixtures to increase the thermal conductivity of the compound material. A common use of a composite target that serves as a good model for this work is in the production of fuel plates for research reactors or plate targets to produce the medical isotope, $^{99}$Mo[9]. The typical composite target uses a mixture of about 50 % by volume of aluminum powder and 50 % powdered uranium-aluminum alloy ($UAl_x$, where x can be 2, 3 or 4), called the fuel meat. In this case, the aluminum serves as the matrix and provides the high thermal conductivity path to remove the heat from the uranium target fuel meat. Figure 1 shows an example of a plate target for producing $^{99}$Mo[9]. The section view on the right has dimensions comparable to what could be used for a target for either proton or electron production of $^{225}$Ac. There is a large database showing a high reliability of the technology of composite UAl$_x$-Al targets. When these targets are used as fuel plates for research reactors there is a long operational period during which the gaseous fission products build up that tends to reduce the thermal conductivity. This is a good model of the buildup of helium and radon during irradiation of a radium target[10].

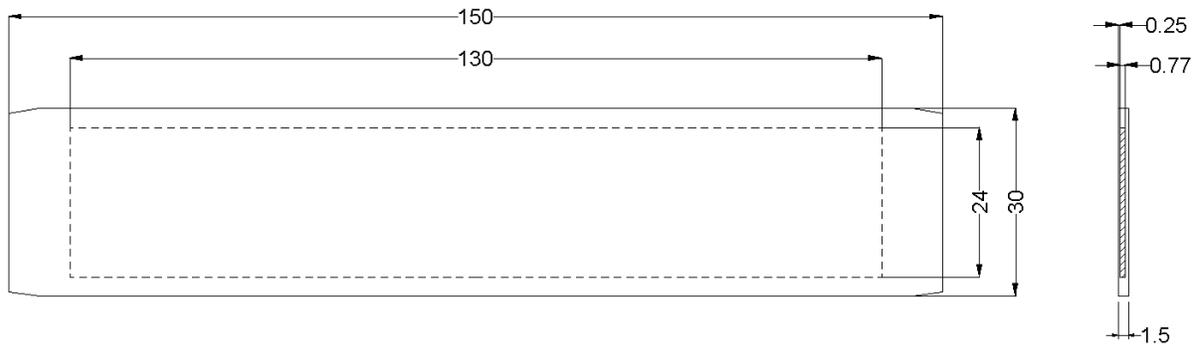

*Figure 1. Plate target using a mixture of aluminum powder and Al$_x$U-Al dispersion mixture for production of $^{99}$Mo with a low enriched uranium (LEU) core[9]. Dimensions are in mm.*

Reference[11] provides details of the fuel composition and thermal hydraulics for use of both high enriched uranium (HEU) and low enriched uranium (LEU) fuel element plates for the BR-2 reactor at Argonne National Laboratory. Table 1 shows some of the key performance data. The maximum heat flux is nearly 500 W/cm$^2$, significantly higher than would be necessary for accelerator production of $^{225}$Ac, as will be shown later. A cooling system producing parameters such as shown in Table 1 can be designed for use with accelerator targets for either electron or proton applications.

*Table 1. Input parameters to establish the operating conditions for BR-2 fuel elements[11].*

| Parameter | Value |
|---|---|
| Maximum Heat Flux | 470 W/cm$^2$ |
| Coolant temperature | 313 K |
| Coolant pressure | 1.24 MPa |
| Coolant velocity | 10.4 m/s |
| Channel thickness | 3.00 mm |

There are several important considerations when using a composite target such as this for proton or electron irradiation:

1) The contribution of the matrix material to the heat transfer from the target material with a low thermal conductivity.
2) The density of the matrix material.
3) The activation of the added material in reactions with protons or electrons.

These are considered in the following sections.

## 3. Properties of Target Materials

The following are some of the relevant physical properties of indium and the potential radium salts that might be used for the composite target. Section 6 considers the activation challenges of indium and radium when they are irradiated with either protons or electron bremsstrahlung photons.

### 3.1. Thermal and Physical Properties of Indium

Indium is a highly ductile metal that is readily available as a powder with a variety of sizes from mm to sub-micron.

Important properties of indium are as follows: melting point, 156.6 °C; boiling point, 2072 °C; thermal conductivity, 82 W/(m-K); density, 7.3 g/cm$^3$; yield strength of approximately 1 MPa.

The low yield strength of indium enables target fabrication with a low die pressure. The high thermal conductivity will contribute to the overall thermal conductivity of the composite target. Indium has a high density of 7.3 g/cm$^3$ that reduces the potential yield of composite targets but it will mix well with radium salts of similar densities and will provide stable operation at reasonable power densities.

### 3.2. Thermal and Physical Properties of Radium Salts

There is much less data about the physical, thermal, and chemical properties of radium. Radium is very radioactive and difficult to handle so many aspects affecting handling are incomplete. Table 2, from reference[12], is a summary of some of the important properties. Note that many entries show no data. $RaCO_3$, $RaCl_2$ and $RaNO_3$ all have molecular weights that are about 20% higher that metallic radium with a density (where known) that is close to the density of metallic radium. Compounds such as $RaCO_3$ are insoluble in water and have greater stability than metallic radium.

An estimate of the density of $RaCO_3$ has been made by considering the ratio of the densities to molecular weights of the other metal carbonates in Group 2: $CaCO_3 = 0.027$; $SrCO_3 = 0.024$; $BaCO_3 = 0.022$. If an extrapolated value of 0.020 is used for $RaCO_3$, then the density would be about 5.7 g/cm$^3$. This value will be used for further calculations with the knowledge that there could be a significant error in the number.

*Table 2 Properties of radium and its salts[12].*

| Property | Radium | Radium Bromide | Radium Carbonate | Radium Chloride | Radium Nitrate | Radium Sulphate |
|---|---|---|---|---|---|---|
| Chemical Formula | Ra | $RaBr_2$ | $RaCO_3$ | $RaCl_2$ | $RaNO_3$ | $RaSO_4$ |
| Molecular Weight | 226.04 | 385.83 | 286.03 | 296.93 | 288.03 | 382.08 |
| Color | Silver-White | White | White | Yellowish-White | No Data | White |
| Physical State | Solid | Solid | Solid | Solid | Solid | Solid |
| Melting Point ($^o$C) | 700 | 728 | No Data | No Data | No Data | No Data |
| Boiling Point ($^o$C) | <1140 | 900 (sublimes) | No Data | No Data | No Data | No Data |
| Density at 20$^o$C | 5 | 5.79 | No Data | No Data | No Data | No Data |
| Solubility in Water at 20$^o$C | Soluble | Soluble | Insoluble | Soluble | Soluble | Insoluble |

### 3.3. Thermal and Physical Properties of Compacted Powders of Indium and CaCO$_3$

As mentioned in the Introduction, we use $CaCO_3$ to estimate the properties of $RaCO_3$. In order to maximize the thermal conductivity of a compacted powder of indium and $CaCO_3$ it is important to construct a target material with a minimum of voids or, equivalently, with low porosity. There are several models that attempt to estimate the density of a compacted powder as a function of the applied pressure. There are also models that estimate the thermal conductivity of a compacted powder. Although these models can provide useful guidance they are not adequate for establishing a target design and estimating its yield.

We have constructed an apparatus, described in the Appendix, that can be used to compact powders and measure the resulting density and thermal conductivity. The measurements have focused on powders of pure indium, pure $CaCO_3$ and a mixture of indium and $CaCO_3$ powders for which the ratio by volume of the compacted material is 40 % indium and 60 % $CaCO_3$.

One of the models suggested by Denny[13] for the dependence of the density on the applied pressure is the Heckel equation which can be written as

$$\ln\left(1/(1-D)\right) = \ln\left(1/(1-D_0)\right) + P/3\sigma_0, \qquad (1)$$

where $P$ is the compaction pressure, $\sigma_0$ is the yield strength, $D_0$ is the relative uncompacted density and $D$ is the relative compacted density. Figure 2 shows a graph of $\ln\left(1/(1-D)\right)$ versus the applied pressure using the measured data for the three powders.

If one considers the data for pure indium, a linear fit predicts a value for the yield strength of 1.3 MPa, in reasonable agreement with values from the literature. The data of most interest for target design is that for the 40/60 In/CaCO$_3$ powder. At an applied pressure of 25 MPa, $\ln\left(1/(1-D)\right)$ has a value of approximately 4, corresponding to a relative density of about 0.98, or a porosity of 2 %. This result indicates that a modest pressure can achieve a density close to the theoretical limit. Although we have studied the behaviour of powders with different ratios of In to CaCO$_3$ we will show later that a ratio of 40/60 is a good compromise for target design.

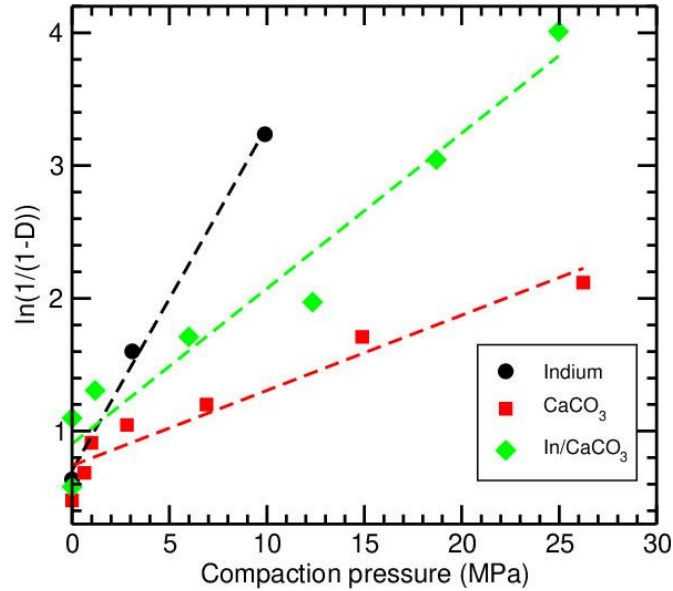

*Figure 2. Measured data (symbols) plotted such that the Heckel equation given by Denny[13] would predict linearity. The relative density is denoted by D and the dashed lines are linear fits to the data. The powder labelled In/CaCO$_3$ is composed of indium and CaCO$_3$ powder in the ratio of 40 to 60 % by volume.*

Composite materials have a wide range of applications and there is an extensive literature on estimating their thermal conductivity. Pietrak and Wiśniewski[14] review many of the models in common use. Some models, such as that of Maxwell-Eucken, are only valid when the fraction of the dispersed phase is less than about 25 %. More sophisticated models require knowledge of the grain shape and size and of the interface effect between the grains. The Effective Medium Model (EMT) as developed by Landauer[15] and discussed by Carson et al[16] is reasonably simple to implement and is expected to apply over a wide range of volume fractions. It can be written as

$$k_m = \big(\left((3\phi_d - 1)k_d\right) + (3(1-\phi_d)-1)k_c$$
$$+ \sqrt{\left((3\phi_d-1)k_d + (3(1-\phi_d)-1)k_c\right)^2 + 8k_c k_d}\,\big)/4, \tag{2}$$

where:

- $k_m$ = thermal conductivity of the mixture,

- $k_c$ = thermal conductivity of the continuous phase,
- $k_d$ = thermal conductivity of the dispersed phase,
- $\varphi_d$ = the volume fraction of the dispersed phase.

Cunningham and Peddicord[17] studied the effect of porosity on the thermal conductivity. For small values of porosity, they found that

$$k_m^{'} = k_m e^{-2.14\phi_v},\qquad(3)$$

where $k_m^{'}$ is the thermal conductivity when the volume fraction of the porosity is $\phi_v$.

In order to estimate $k_m$, the thermal conductivities of the continuous and dispersed phase must be known. For indium, $k_c$ is 82 W/m-K. The thermal conductivity is unknown for any of the radium salts. However, the thermal conductivity of $CaCO_3$ is approximately 2.3 W/m-K[18] and may be indicative of what to expect for $RaCO_3$. Figure 3 shows the prediction of the EMT model for the thermal conductivity as a function of the volume fraction of $CaCO_3$. Also shown is the measured thermal conductivity for a volume fraction of 0.6. The agreement with the model prediction is quite satisfactory. The measured value of 13.9 W/m-K will be used to estimate the thermal behaviour of various target designs.

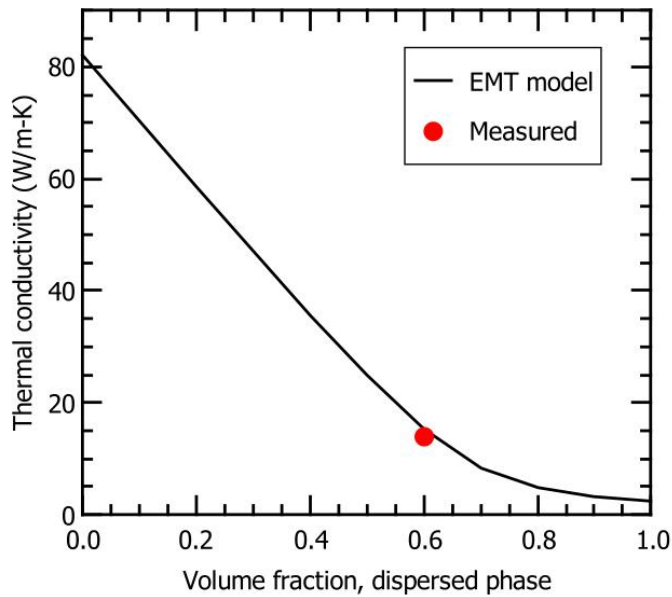

*Figure 3 EMT model estimate of the thermal conductivity of a mixture of indium and CaCO₃ as a function of the volume fraction of CaCO₃. The measured value for a volume fraction of 0.6 is also shown.*

## 4. Target Technology for Proton Irradiation

Figure 4 compares the cross section for the reaction $^{225}Ra(p,2n)^{226}Ac$ as calculated by the Monte Carlo code Fluka[19]–[21] and as measured by Apostolidis et al [22] and Horn et al[23]. The measured cross sections tend to be smaller than the one used by Fluka but they all indicate a peak in the cross section between about 15 and 17 MeV. To optimize the reaction yield and the power budget in the target, the proton energy should be chosen to fall within that energy range. The input energy will be determined by the proton energy from the cyclotron and the energy losses in the target chamber window and other absorbers in the beam line such as a double Havar window. The exit energy will be determined by the energy loss in the target material. The two vertical lines show estimates of the input and the exit energy for a composite target of 40 % indium and 60 % RaCO₃, irradiated with a 24 MeV cyclotron. This target is described below.

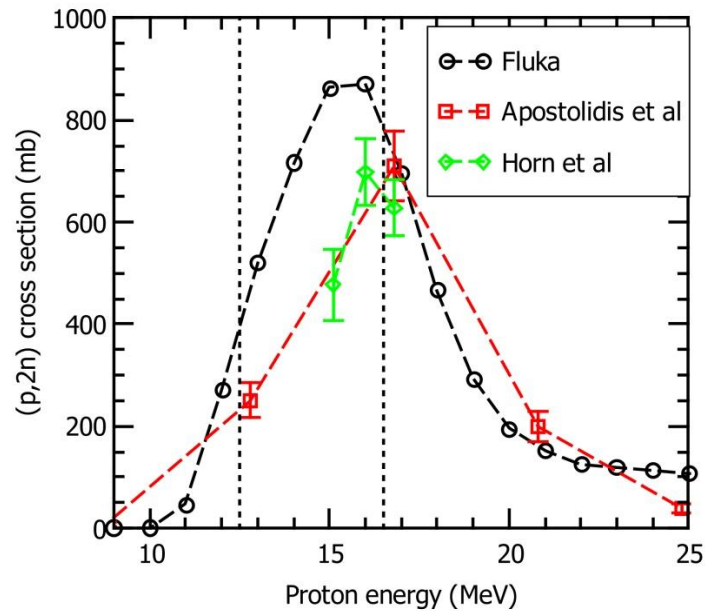

*Figure 4. Comparison of the cross sections for the reaction $^{225}Ra(p,2n)^{226}Ac$ as calculated by the Monte Carlo code Fluka (black circles), as measured by Apostolidis et al[22] (red squares) and as measured by Horn et al[23] (green diamonds) The vertical dashed lines indicate the approximate proton energy range within a 40/60 In/RaCO₃ target for the geometry shown in Figure 5.*

Figure 5 shows a schematic of the setup used to irradiate the target. The cyclotron vacuum system is terminated with a double Havar window that is cooled by helium. This approach isolates the cyclotron from the target material and should provide a high margin of safety with respect to protecting the cyclotron from possible alpha contamination if a target is damaged. Steyn et al[24] has successfully used 75 µm for the upstream window facing the vacuum system and 50 µm for the downstream window. They have also used an aluminum target chamber with an aluminum window machined as part of the chamber[25]. The window could also be mounted using spring loaded and coated metal O-rings to enable window replacement.

It would be possible to use the target chamber as part of the cyclotron vacuum system and eliminate the Havar windows and save about 1.5 MeV energy loss in the windows. There will still be the target encapsulation and the target chamber window to act as barriers between the radium target and the cyclotron vacuum system. It is likely that all windows and encapsulations will fail at some time, but it is much less likely that all fail at the same time. Use of a target with a separated water system and a separate cyclotron vacuum window should further reduce the likelihood of a major failure that contaminates the cyclotron with alpha contamination. These issues need to be addressed as part of the general safety analysis of the project.

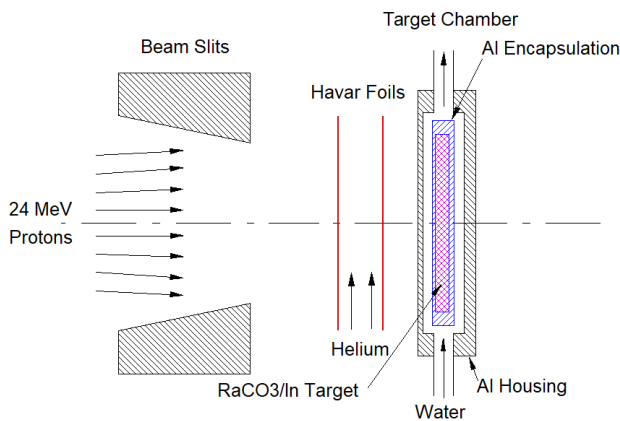

*Figure 5. Schematic of an isotope target using the composite mixture of RaCO₃ and indium.*

Figure 5 shows a proton energy of 24 MeV. This is a common cyclotron energy, and it is about the minimum energy that could be used for the target shown in Figure 6. A proton energy between 24 and about 30 MeV would enable greater flexibility in the hardware design.

The following shows suggested steps to fabricate a composite target that uses 0.5 g of radium – other approaches are possible:

1) The dimensions of the target for cyclotron use are chosen to contain 0.5 g of radium in a 2.5 cm diameter by 0.38 mm deep target, with a volume of about $0.19$ cm$^3$. The Appendix shows that a mixture of 40 % indium and 60 % CaCO$_3$ has a density of about 40 % of the solid density and it should be reasonable to reach a compacted density of over 95 % of solid density. This suggests an initial volume is about $0.47$ cm$^3$, and an initial thickness of 0.95 mm.
2) $0.19$ cm$^3$ of indium powder and $0.28$ cm$^3$ of RaCO$_3$ powder are mixed to a reasonable uniformity. There are no requirements for a high level of uniformity,
3) The powder is added to a target blank as shown in Figure 6. The target blank has an extended rim to contain the extra volume of powder,
4) The target material can be compressed with modest die pressure. The results from the Appendix showed that the target would reach near theoretical density with a compaction pressure of 25 MPa.
5) The target lid is then attached using either cold welding[24] or with a laser weld. The thin rim collapses into the small cutout in the base when the lid is pressed into place.
6) As an alternate approach, both steps of pressing the mixture into place and forming the cold weld could be done in one step.
7) There is additional space where the rim collapses that can serve to collect helium during irradiation.

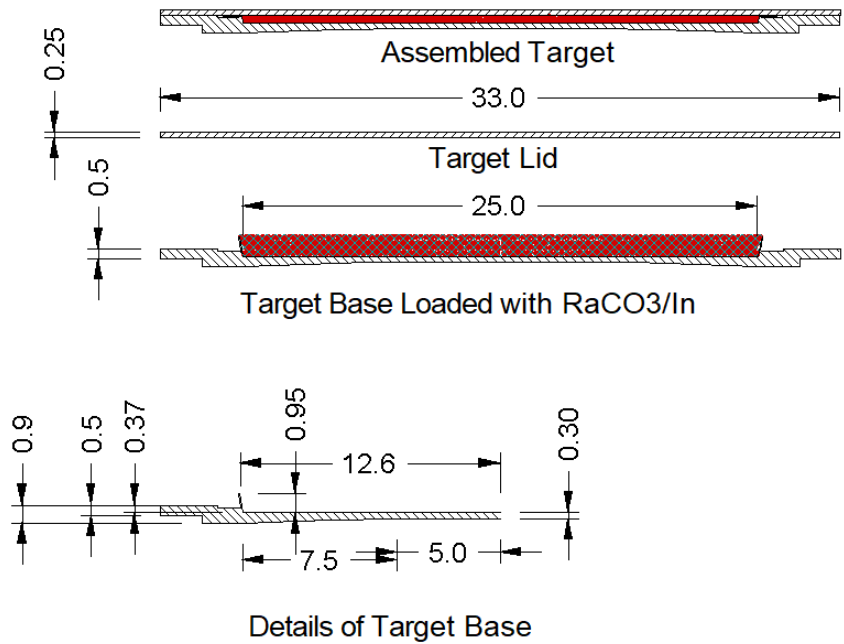

*Figure 6 Fabrication of a cyclotron target using a composite of indium and RaCO₃. The bottom provides some details of the base showing the increase in thickness after 5 mm radius to provide a thin shell at the inner section to reduce power loss and increased strength at outer radius. The middle shows the separate components with the target material above the top, contained by a thin lip and the top shows the assembled target. Dimensions are in mm.*

Table 3 and Table 4 show the actual thicknesses of the components of a detailed layout of the target and chamber. Fluka was used to calculate the power dissipated in each component. The proton beam was assumed to have an energy of 24 MeV, a Gaussian profile with a full width half maximum (FWHM) of 11 mm and a beam power of 3 kW. Table 3 refers to the materials upstream of the face of the target material while Table 4 refers to the target material and the downstream components.

*Table 3. Target description for the components upstream of the target face. The beam energy is 24 MeV and current is 125 µA. The power in each component was calculated using Fluka.*

| Layer Description | Thickness | Material | Power (W) |
|---|---|---|---|
| Beamline Window # 1 | 75 µm | Havar | 115 |
| Coolant | 10 mm | He | 0.05 |
| Beamline Window # 2 | 50 µm | Havar | 79 |
| Air | 4 mm | - | 1.27 |
| Target Chamber Entrance Window | 0.6 mm | Aluminum | 387 |
| Water | 0.5 mm | Water | 168 |
| Target Entrance Window | 0.25 mm | Aluminum | 182 |
| Total | | | 932 |

*Table 4. Target description for a target of 40% In and 60% RaCO₃, and the downstream components. The beam energy is 24 MeV and current is 125 µA. The power in each component was calculated using Fluka.*

| Layer Description | Thickness | Material | Power (W) |
|---|---|---|---|
| Target (In/RaCO3) | 0.38 mm | 40% Indium +60% RaCO3 | 491 |
| Target Exit Window | 0.3 mm | Aluminum | 306 |
| Water | 1 mm | Water | 762 |
| Target Chamber Exit Window | 2 mm | Aluminum | 466 |
| Total | | | 2025 |
| Total for Target | | | 2957 |

### 4.1. Yield Estimates

Table 5 shows the results of calculations with Fluka for the yield of $^{225}$Ac at the end of bombardment (EOB) from a 10-day irradiation (about one half-life) of the target described in Table 3 and Table 4.

*Table 5. The yield (EOB) of $^{225}$Ac produced by irradiating the targets described in Tables 4 and 5 for 10 days. The proton beam had an energy of 24 MeV, a Gaussian beam profile with a full-width at half-maximum (FWHM) of 11 mm and a beam current of 125 µA. The Monte Carlo statistical uncertainty is given in column 3.*

| # 225Ac/p | Activity | u_c |
|---|---|---|
| | GBq (Ci) | GBq (Ci) |
| $2.03 \times 10^{-4}$ | 79.6 (2.15) | 0.5 (0.01) |

In 2018, Robertson et al[26] estimated the annual requirement for $^{225}$Ac at about 185 GBq (5 Ci) per year. A single facility, with the production rate given in Table 5, can easily meet this requirement.

### 4.2. Thermal Management

We used the Elmer software package[27] to calculate the thermal behavior of the proposed proton target. Elmer is based on the finite element method (FEM) and requires as input the power density distribution throughout the target assembly. This was calculated using Fluka. The thermal conductivity of aluminum and In/RaCO$_3$ were taken to be 167 and 14 W/m-K, respectively.

An important consideration is the thermal conductance, $h$, at the interfaces between the target material and the aluminum wall and between the aluminum wall and the water. There is an extensive literature on the thermal conductance between metal surfaces because of its industrial importance and the book by Madhusudana[28] discusses measurements and models. The conductance depends on the type of materials in contact, their surface finish, the applied pressure and any gas present in the interface. Typical values (see Reference [29]) can range from as low as 0.02 W/cm$^2$-K with no gas in the interface to as high as 4.5 W/cm$^2$-K. In many cases, $h$ can be enhanced by adding some material to the interface, such as thermally conducting paste, helium gas or indium foil[30]. Ainscough[31] has reviewed the data on $h$ for reactor fuel rods and values range from 1 to 3 W/cm$^2$-K. Because we have a target material that already contains a large fraction of indium it is not obvious how to estimate $h$. The Appendix reports a measured value of 3.5 W/cm$^2$-K and this value will be used in the thermal model.

The thermal conductance at the interface between aluminum and water has been measured[32]and calculated[33]. Assuming laminar flow of the water over the aluminum surface the value of $h$ is typically about 6000 W/cm$^2$-K. Because the heat fluxes we encounter is this work is typically less than 500 W/cm$^2$ the temperature step at the aluminum-water interface can be ignored.

The temperature was calculated throughout the target but Figure 7 shows the temperature along the central beam axis and along the central radius of the target. It was assumed that water flowing around the target can keep the outer aluminum wall at 30 °C. The beam power dissipated in the target material raises its temperature above that of the aluminum by about 13 °C but the thermal impedance of the gap leads to an additional increase of about 52 °C.

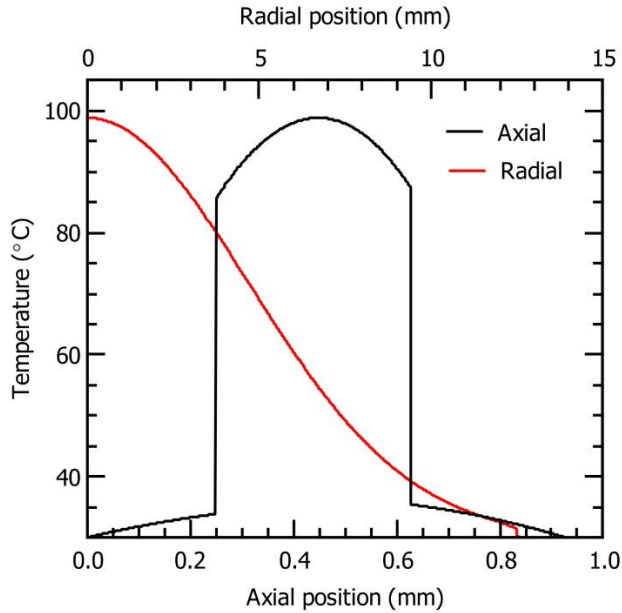

*Figure 7. Elmer FEM calculation of temperature profiles in the proton target assembly. The water flowing around the aluminum target holder was assumed to hold the outer wall of the aluminum at 30 °C.*

Figure 8 shows the heat fluxes on the four principal interfaces of the target assembly. The maximum heat flux at the aluminum-water interface is about 400 W/cm$^2$ and this power must be carried away by the water flowing around the target. There is an extensive literature[34] on heat removal using water flowing at a high pressure and a large flow rate. However, the data shown in Table 1 indicates that a heat flux of 470 W/cm$^2$ is achieved for the BR-2 fuel elements.

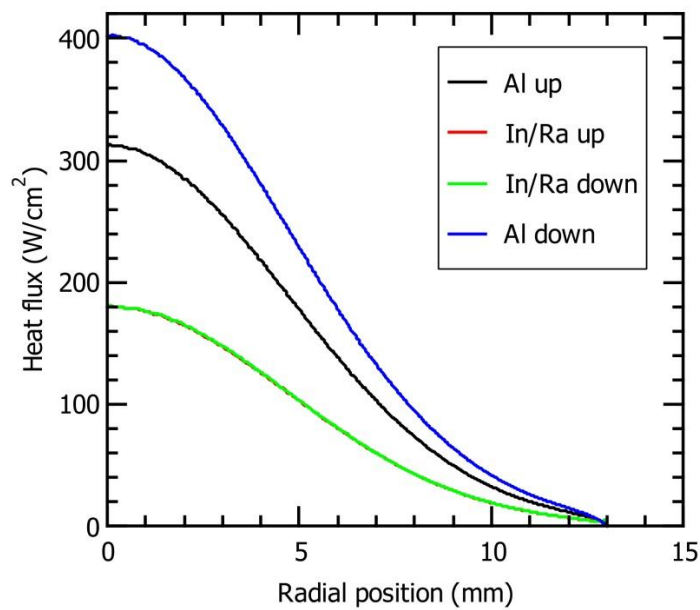

*Figure 8. Elmer calculation of the heat flux on various surfaces of the target assembly Note that the red and green curves are almost identical.*

There are also practical techniques to reduce the maximum heat flux, should that prove necessary. The present model assumes the proton beam has a Gaussian profile. This leads to a high heat flux near the central axis. Figure 9 compares different beam profiles and shows that if one sweeps the proton beam in a small circular arc a more uniform distribution of protons on the target face can be achieved.

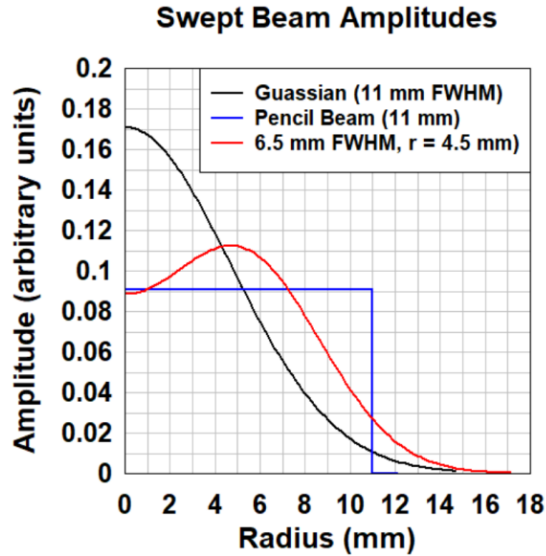

*Figure 9. Comparison between a uniform, 11-mm radius (pencil) beam, a Gaussian beam profile of 11 mm (FWHM) and a scanned Gaussian beam of 6.5 mm (FWHM) and 4.5 mm scan radius. The area of each profile is unity.*

## 5. Target Technology for Electron Irradiation

### 5.1. Target Details

Figure 10 shows a schematic of a setup using the same basic components as in Figure 5 with the addition of three 1- mm thick tantalum plates that act as a converter target to convert the energy of the electrons into bremsstrahlung that produce the reactions. There is also a thin window assembly at the end of the accelerator beam line, as shown in Figure 5.

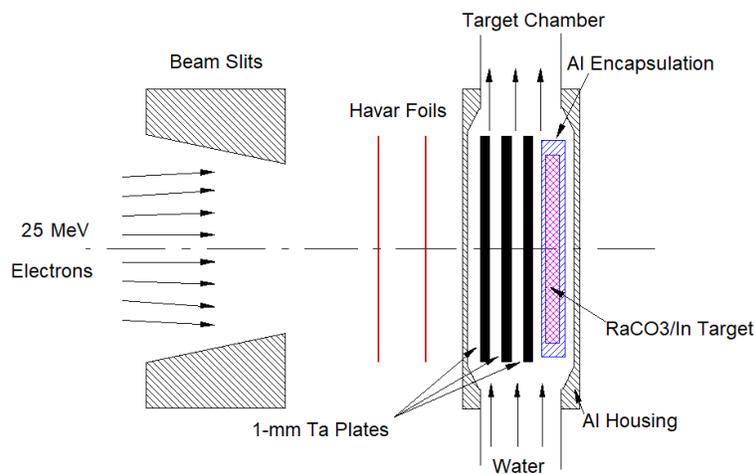

*Figure 10. Schematic of a setup to irradiate a composite RaCO₃ and indium target with electrons.*

Figure 11 shows the details of a potential electron accelerator target designed using the same procedures as outlined in Section 4. The dimensions of the target material are 10 mm diameter by 1.17 mm deep with a volume 0.093 cm$^3$. The thickness is chosen to produce 0.25 g of radium in each target for comparison with the radium metal targets described in Reference [10]. This leads to a volume of 0.056 cm$^3$ of RaCO$_3$. Using the density of 5.7 g/cm$^3$ (see Section 3.2) produces a mass of 0.319 g of RaCO$_3$ and a mass of radium of 0.319 x 226/286 = 0.25 g. There will also be 0.037 cm$^3$ of indium at a density of 7.3 g/cm$^3$ = 0.27 g, producing a total mass of 0.59 g per 0.093 cm$^3$ = 6.30 g/cm$^3$. Measurements described in the Appendix and summarized in Figure 2 show that near theoretical density can be achieved with modest die pressures of the order of about 25 MPa. There is also an annular channel at the outer radius of the target material to enable collection of the helium produced during irradiation.

Using an uncompacted density of about 40% leads to an initial volume of 0.23 cm$^3$ and a depth of 0.30 cm for the lip of the target base before assembly.

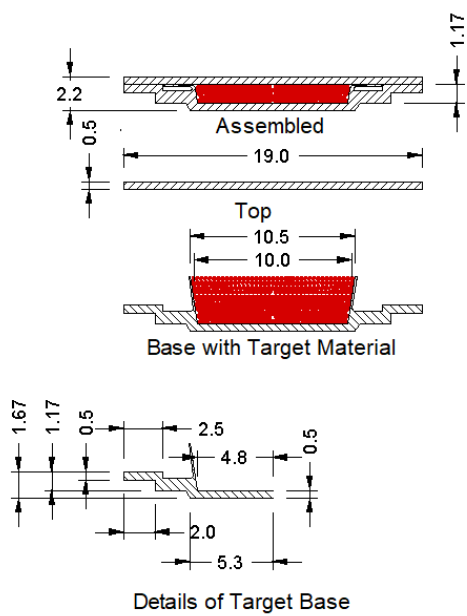

*Figure 11. Details of a potential target for electron accelerator use Dimensions are in mm.*

Figure 12 shows a layout of four isotope targets of the dimensions shown in Figure 11. The upper half of Figure 12 shows the configuration from Reference [10] with three 1-mm tantalum converter targets and four radium metal targets of dimensions 10 mm diameter by 0.63 mm deep with a mass of 0.25 g of radium in each target, and an electron beam of 7 mm FWHM. The aluminum encapsulation is 0.5 mm thick at front and back. The lower half has the same converter target configuration and four In/RaCO$_3$ targets as shown in Figure 11 with the same spacing between targets (1.0 mm) and the windows at the end of the accelerator. Other configurations for different target thicknesses and energies should scale reasonably close to the calculations shown in Reference [10].

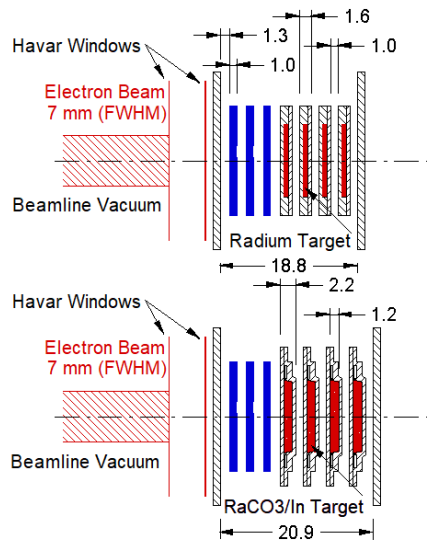

*Figure 12. Details of a complete target assembly for electron irradiation Dimensions are in mm.*

### 5.2. Monte Carlo Calculations of $^{225}$Ra Yields

Table 6 shows a summary of the results of Fluka calculations for the yield at End of Bombardment (EOB) for several configurations, including the results for four 0.25 g metal radium target for comparison to Reference [10]. The yields of the composite target are modestly lower than the pure metal radium of the same mass, as would be expected. The loss in yield is a combination of the increase in absorption in the additional target material, but mostly from the loss of photons because of the additional distance between the isotope targets and converter targets produced by the larger physical dimensions of the composite targets. There are four sets of data reported in Table 7. The following is a summary of those calculations:

1) Four 0.25 g radium metal targets.
2) The third column is the reference target using the composite targets shown in Figure 11 and Figure 12. The yield is only reduced by about 8 % compared to the metal radium target case.
3) The fourth column shows the results with four 1-mm tantalum plates in the converter. The radium target geometry remains the same as in column 3. The yield is about 74 % of the metal targets because of the higher absorption of photons in the additional tantalum plate and because of geometric effects of increasing the distance between the converter and isotope target.
4) The fifth column shows the results with five 1-mm tantalum plates in the converter. The radium target geometry remains the same as in column 3. The yield is about 60 % of the metal targets because of the higher absorption of photons in the additional tantalum plate and because of geometric effects of increasing the distance between the converter and isotope target.

Although electron irradiation produces $^{225}$Ra and not $^{225}$Ac directly, it turns out than three milkings of $^{225}$Ra yields approximately the same activity as the starting activity of $^{225}$Ra[10]. Thus, the $^{225}$Ra yields reported in Table 6 can be interpreted as approximate $^{225}$Ac yields.

*Table 6. Summary of the calculated yields of $^{225}$Ra for the two configurations in Figure 12 for a 10-day irradiation at 25 MeV and 20 kW with a beam width of 7 mm FWHM and for two runs with additional tantalum plates in the converter target. The statistical uncertainty of the Monte Carlo calculations is about 0.5 %.*

| Targets | Yield, EOB GBq (Ci) Pure Ra with 0.25 g Ra 3 Ta Plates | Yield, EOB GBq (Ci) 40/60 In/RaCO3 with 0.25 g Ra 3 Ta Plates | Yield, EOB GBq (Ci) 40/60 In/RaCO3 with 0.25 g Ra 4 Ta Plates | Yield, EOB GBq (Ci) 40/60 In/RaCO3 with 0.25 g Ra 5 Ta Plates |
|---|---|---|---|---|
| 1 | 33.7 (0.91) | 33.0 (0.89) | 26.1 (0.70) | 20.5 (0.56) |
| 2 | 27.0 (0.73) | 25.3 (0.68) | 20.3 (0.55) | 16.1 (0.44) |
| 3 | 22.2 (0.60) | 19.7 (0.53) | 15.9 (0.43) | 13.0 (0.35) |
| 4 | 18.5 (0.50) | 15.8 (0.43) | 13.0 (0.35) | 10.7 (0.29) |
| Total | 101.4 (2.74) | 93.8 (2.53) | 75.3 (2.03) | 60.3 (1.64) |
| % of Col. 1 | | 92 | 74 | 60 |

A separate calculation showed that the effect of the Havar windows was a decrease of from 3 to 4 % compared to no windows, a small effect. The yield of the composite targets (Column 3) is close to that of the metal radium targets but the next section shows that the power dissipated in the first radium target is too high for practical use. There are several ways to reduce the power dissipated in the radium targets. A common method is to use additional tantalum plates in the converter target, to absorb a larger fraction of the incident electrons. The collision range of 25 MeV electrons is 10.85 g/cm$^2$ and the range in tantalum (density of 16.7 g/cm$^3$) is 6.5 mm. For 25 MeV electrons, about 50 % of the energy is dissipated stopping the electrons and 50 % as bremsstrahlung radiation that produces the photonuclear reactions and has a greater range than the electrons. The next two columns show the yields for the same four radium targets with one and two additional 1-mm thick tantalum converter plates. Five mm of tantalum plus the titanium window and water reduces the power from the electrons and most of the energy is from absorbing bremsstrahlung.

### 5.3. Monte Carlo Calculations of the Power Deposition

Table 7 shows the values calculated by FLUKA for the power deposited in each element of the target assembly for the four cases noted in Table 6. The calculations are for a beam power of 20 kW, a beam profile of 7 mm FWHM. Note the rapid decrease in the power in the first In/RaCO$_3$ target material, denoted by Ra-1, as the number of Ta plates increases.

*Table 7 Power deposited in each element of the target assembly for a beam power of 20 kW and beam energy of 25 MeV at a beam width of 7 mm FWHM. The statistical uncertainty of the Monte Carlo calculations is typically less than 0.1 %.*

| Element | Pure Ra with 0.25 g Ra (W) | 40/60 In/RaCO3 with 0.25 g Ra (W) | 40/60 In/RaCO3 with 0.25 g Ra (W) | 40/60 In/RaCO3 with 0.25 g Ra (W) |
|---|---|---|---|---|
| # of Ta Plates | 3 | 3 | 4 | 5 |
| Window # 1 Havar | 67.7 | 67.7 | 67.7 | 67.8 |
| He | 0.05 | 0.05 | 0.05 | 0.05 |
| Window # 2 Havar | 46.7 | 46.7 | 46.8 | 46.8 |
| Air | 0.97 | 0.97 | 0.97 | 0.97 |
| Window #1 (Ti) | 576 | 576 | 578 | 578.3 |
| Window #2 (Ti) | 197 | 148 | 89.5 | 65.1 |
| Ta-1 | 2,098 | 2,099 | 2,101 | 2,101 |
| Ta-2 | 2,554 | 2,555 | 2,562 | 2,563 |
| Ta-3 | 2,044 | 2,045 | 2,119 | 2,119 |
| Ta-4 | | | 1,230 | 1,263 |
| Ta-5 | | | | 608.9 |
| Ra-1 | 139.5 | 329.7 | 161.8 | 76.3 |
| Ra-2 | 84.5 | 172.1 | 82.8 | 43.6 |
| Ra-3 | 53.2 | 93.8 | 47.3 | 29.0 |
| Ra-4 | 34.1 | 53.4 | 30.4 | 21.6 |
| Al-1 | 159.2 | 158.9 | 76.1 | 34.5 |
| Al-2 | 95.8 | 82.1 | 38.1 | 19.2 |
| Al-3 | 59.6 | 44.2 | 21.5 | 12.8 |
| Al-4 | 37.8 | 25.0 | 13.8 | 9.67 |
| Water | 2,210 | 2,286 | 2,146 | 2,104 |
| Total | 10,460 | 10,784 | 11,413 | 11,766 |
| Escape | 9,541 | 9,216 | 8,586 | 8,234 |

Using the same approach as discussed in Section 4.2, the Elmer code was used to calculate the temperature profiles and heat flux for the first In/RaCO$_3$ target when four Ta plates are used. The results are shown in Figure 13 and Figure 14. The beam power in the target material raises its temperature above that of the aluminum wall by about 31 °C. The thermal impedance of the interface leads to a further increase of about 46 °C.

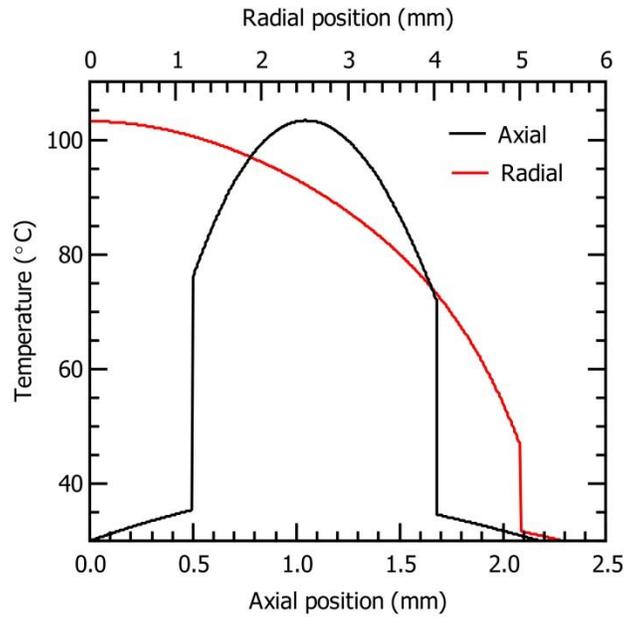

*Figure 13. Elmer FEM calculation of temperature profiles in the electron target assembly. This is for the first In/RaCO₃ target when four Ta plates are used. The water flowing around the aluminum target holder was assumed to hold the outer wall of the aluminum at 30 °C.*

Figure 14 is equivalent to Figure 8 but for the first In/RaCO$_3$ target assembly shown in the bottom section of Figure 12. In this case, the maximum heat flux that must be carried away by the water is about 200 W/cm$^2$.

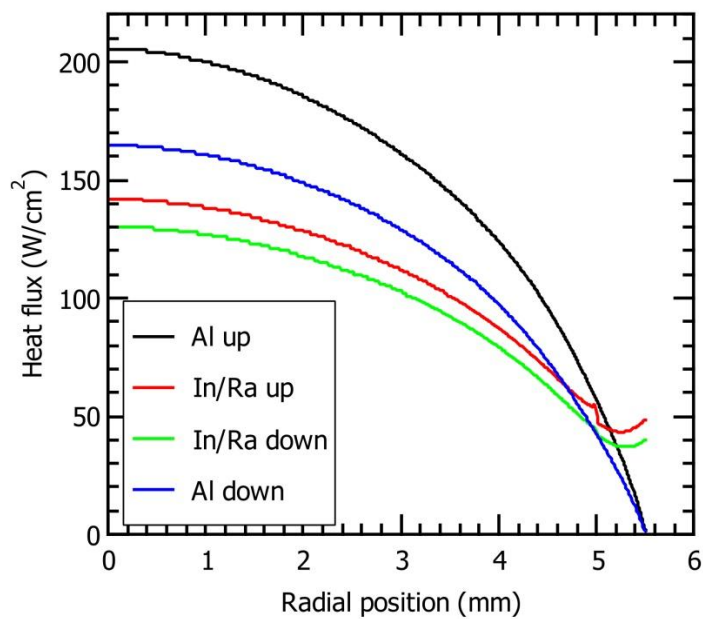

*Figure 14. Elmer calculation of the heat flux on various surfaces of the target assembly when four Ta plates are used.*

## 6. Nuclear Properties

This section considers the activation products of the two materials after they have been irradiated by protons or electron bremsstrahlung photons. These products may affect the challenges of handling the target post irradiation.

### 6.1. Indium

Indium is mostly (96 %) [115]In with 4 % as [113]In. A review of the Chart of the Nuclides shows that for proton irradiations, all of the (p,xn) reactions on [115]In lead to stable isotopes and for the 4 % [113]In isotope there will be a modest yield of the 115 d half-life [113]Sn that decays mainly via a 392 keV photon (65 %). During a 5- to 10-day irradiation this will only reach a few percent of saturation activity and should not pose a serious handling problem.

Table 8 shows the results of a Fluka calculation for a 150 µA beam of protons incident on a 0.4 mm thick natural indium target, typical of an expected irradiation of the composite target.

*Table 8 The results of a Fluka calculation of a 5-day irradiation with a 150 µA beam of 18 MeV protons incident on a 0.4 mm thick natural indium target. In column 2, the statistical uncertainty in the last digit is given in parentheses.*

| Isotope | Total # Produced | Half Life | Activity (EOB) GBq (mCi) |
|---------|------------------|-----------|--------------------------|
| $^{109}$Cd | $7.5(8) \times 10^{13}$ | 462 d | 0.0013 (0.035) |
| $^{110}$Cd | $1.7(1) \times 10^{14}$ | stable | |
| $^{111}$Cd | $7.6(3) \times 10^{14}$ | stable | |
| $^{112}$Cd | $7.9(3) \times 10^{14}$ | stable | |
| $^{112}$In | $1.3(1) \times 10^{14}$ | 15 m | 0.290 (7.85) |
| $^{113}$In | $1.09(3) \times 10^{15}$ | stable | |
| $^{114}$In | $1.77(5) \times 10^{15}$ | 72 s | 4.09 (111) |
| $^{115}$In | $2.06(1) \times 10^{16}$ | stable | |
| $^{112}$Sn | $2.124 \times 10^{16}$ | stable | |
| $^{113}$Sn | $1.05(1) \times 10^{16}$ | 115 d | 0.722.(19.5) |
| $^{114}$Sn | $6.391(6) \times 10^{17}$ | stable | |
| $^{115}$Sn | $9.06(3) \times 10^{16}$ | stable | |

The isotopes labelled in red are radioactive. At the end of a five-day irradiation, most of the activity (about 0.1 Ci) is from $^{114}$In. However, its half life is short (72 s) so is unlikely to be a hazard.

For photon irradiations, all of the (γ,xp) reactions on both indium isotopes lead to stable isotopes. The dominant reaction is $^{115}$In(γ,n)$^{114}$In that decays rapidly and for the 4 % $^{113}$In isotope there will be a modest yield of the 2.8 d half-life $^{111}$In that decays mainly via 171 and 245 keV photons with probabilities of 91 % and 94 %, respectively.

Table 9 gives the results of a Fluka calculation for a 10-day irradiation with a 25 MeV, 20 kW beam of electrons incident on 0.5-mm natural indium target.

*Table 9 The results of a Fluka calculation of a 10-day irradiation with a 25 MeV, 20 kW beam of electrons incident on ½-mm natural indium target. The target configuration is given in Section 5.1 with indium replacing the $1^{st}$ radium target. In column 2, the statistical uncertainty in the last digit is given in parentheses.*

| Isotope | Total # Produced | Half Life | Activity (EOB) GBq (mCi) |
|---------|---------|---------|---------|
| $^{111}$In | $1.6(1) \times 10^{14}$ | 2.8 d | 0.17 (4.6) |
| $^{111}$Ag | $9(1) \times 10^{12}$ | 7.45 d | 0.006 (0.2) |
| $^{112}$In | $8.38(8) \times 10^{15}$ | 14.8 min | 9.69 (262) |
| $^{112}$Cd | $9(1) \times 10^{13}$ | stable | |
| $^{113}$In | $1.10(1) \times 10^{16}$ | stable | |
| $^{113}$Cd | $3(2) \times 10^{12}$ | stable[1] | |
| $^{114}$In | $2.515(5) \times 10^{17}$ | 71.9 s | 291 (7,870) |
| $^{114}$Cd | $1.98(5) \times 10^{15}$ | stable | |
| $^{115}$In | $3.7(5) \times 10^{13}$ | stable[1] | |

[1] Half-lives of $^{113}$Cd and $^{115}$In: $8.04 \times 10^{15}$ and $4.40 \times 10^{14}$ years.

The isotopes labelled in red are radioactive. As was the case for proton irradiation, most of the activity (about 8 Ci) is from $^{114}$In. However, its half life is short (71.9 s) so is unlikely to be a hazard.

If the radium salt contains carbon, oxygen, or nitrogen, the reactions produced by either protons or electrons lead to products that rapidly decay. If $RaCl_2$ is used there are reaction products that may need to be considered.

### 6.2. Radium

It was noted earlier that $^{226}$Ra has a high specific activity for a target material. One Ci (37 GBq) was originally based on the activity of one gram of $^{226}$Ra. One of the daughter products in equilibrium with the $^{226}$Ra is $^{214}$Bi that decays with high-energy photons that are highly penetrating. Production of several Ci of $^{225}$Ac will require the use of target masses of at least hundreds of mg of $^{226}$Ra. During production of multi-curie yields of $^{225}$Ac over irradiation periods of five to ten days, there are more isotopes that decay with high-energy photons, especially $^{208}$Tl that is a decay product of the $^{224}$Ra produced by either proton or electron induced reactions. Table 10 shows the expected gamma-radiation field from one curie of $^{226}$Ra and from the other likely products from either of the two accelerator production methods using radium targets. The calculations were done with the shielding code Microshield[35] and assume secular equilibrium with all decay products. The isotopes produced by either accelerator method may not be in secular equilibrium at EOB and may produce significantly lower gamma radiation fields than calculated. Even using 10 cm of lead equivalent for the hot cell the gamma fields at a typical working distance from the source are significant.

*Table 10 Gamma radiation fields at 50 cm from a one-curie (37 GBq) source of $^{226}$Ra, $^{225}$Ra and $^{224}$Ra. These Microshield[35] calculations assume secular equilibrium of all decay products.*

| Isotope[1] | Half Life | Unshielded mSv/h (mR/h) | 5 cm of Pb mSv/h (mR/h) | 10 cm of Pb mSv/h (mR/h) |
|---|---|---|---|---|
| Ra-226 | 1600 y | 9.10 (910) | 1.50 (150) | 0.022 (2.2) |
| Ra-225 | 14.9 d | 2.68 (268) | 0.048 (4.8) | |
| Ra-224 | 3.6 d | 8.24 (824) | 1.42 (142) | 0.033 (3.3) |

## 7. Practical Considerations

Radium carbonate (or another suitable radium salt) can be prepared as a powder in a laboratory with the necessary facilities for handling radium. Indium powder is available commercially[36] in a range of sizes. One will choose the indium powder grain size to roughly match that of the RaCO$_3$. As described in the Appendix, suitable amounts of indium and RaCO$_3$ powders will be weighed to establish a volume ratio of 40 % to 60 % in the compacted material. Small commercial shakers are available[8] that can be used to homogenize the powder. The resulting mixture can be pressed using a commercial press[37]. The die used to press the powder will be chosen to match the proposed target diameter. Aluminum crucibles are available[38] in approximately the correct size as well as a press for cold welding the lid.

## 8. Summary

We have shown that indium is a suitable matrix material for creating a composite target with a radium salt. Using CaCO$_3$ as an analog for RaCO$_3$ we have constructed an apparatus that could be used to press the powders. Although different volume ratios of indium to CaCO$_3$ were tested a ratio of 40 indium to 60 % CaCO$_3$ (or RaCO$_3$) was considered a reasonable compromise. The measured thermal conductivity for the mixture is in reasonable agreement with the Heckel model.

We used the Fluka Monte Carlo code to calculate the expected yields of $^{225}$Ac or $^{225}$Ra for various target configurations. Figure 4 indicates that the proton cross section used by Fluka is significantly larger than measured cross sections so our predicted proton yields may be too large. On the other hand, the photonuclear data presented in Reference [10] indicates that the Fluka predictions for an electron beam may be too low.

The Elmer finite element code was used to calculate the thermal properties of the targets. There are two constraints that must be considered. The maximum temperature in the target material cannot exceed the melting point of indium and the heat flux from the surface of the target must be low enough that it can be transported away by flowing water. We show that the proposed target geometries and irradiation parameters can satisfy these requirements. We note that the thermal resistance at the interfaces between the target material and the aluminum holder has a major impact on the target temperature.

Using either a 24 MeV proton beam or a 25 MeV electron beam, it is practical to produce approximately 74 GBq (2 Ci) of $^{225}$Ac using a 10-day irradiation. In 2108, Robertson et al[26] estimated the annual demand for $^{225}$Ac at 185 GBq (5 Ci). A single irradiation facility could meet this requirement.

## Appendix. Heat Transfer Measurements of Composite Targets of Compacted Indium and CaCO₃ Powders

### A1. Apparatus

Experiments have been conducted to demonstrate the effectiveness of composite targets of indium powder and CaCO$_3$, one of the salts in the Group 2 elements, to transfer reasonable power densities such as might be used for a radium salt target. The measurements have been made on the apparatus shown in Figure 15 and Figure 16. It consists of a top piece made from 6061 aluminum alloy with a thermal conductivity of approximately[39] 160 W/m-K. There is a hole in one side for insertion of a thermocouple and it is machined to a diameter of 1.9 cm to produce an area of about 2.8 cm$^2$. The top is machined to accept an electric heater. The insert can hold up to 4.0 mm of material to be tested. It also has a hole in one side to accept a thermocouple. The insert sits on an aluminum base that also has a hole to accept a thermocouple. The right-hand side shows the apparatus assembled with target material in the insert. Two ¼-20 nuts threaded on inserts from the base, are used to provide compaction of the polyimide insulator and the target sample under test. The force ,$F$, exerted by a bolt can be related to the torque, $T$, on the nut by[40]

$$F = T/Kd, \qquad (4)$$

Where $d$ is the bolt diameter and $K$ is a constant with a value of about 0.2 for typical bolts. For a torque of 1 N-m the force exerted by the bolt will be about 790 N. The pressure applied to the sample by the two bolts will be about 5.6 MPa. The maximum torque applied was about 5 N-m, leading to a compaction pressure of about 28 MPa.

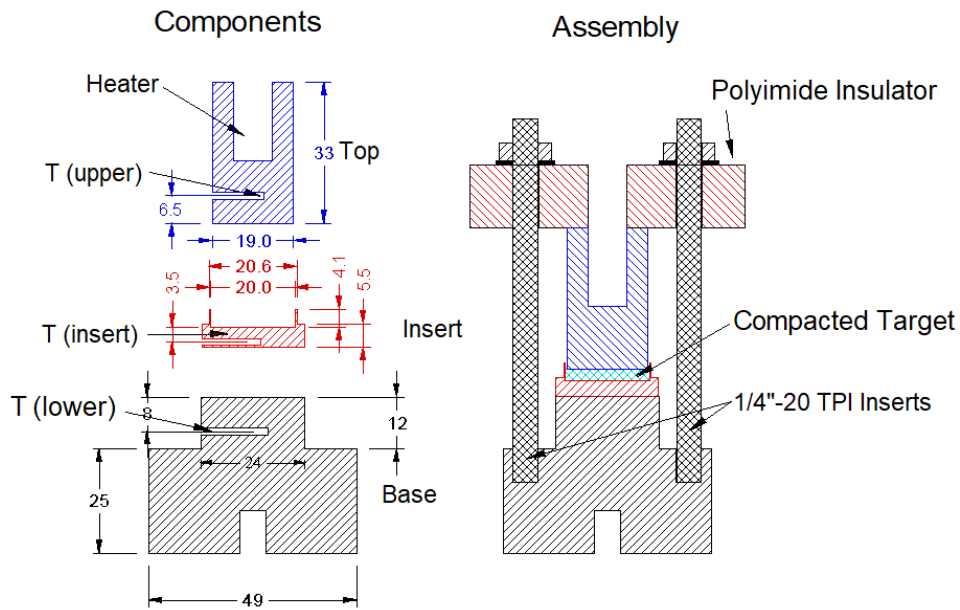

*Figure 15. Details of the measurement apparatus. All of the components on the left hand side are constructed from aluminum. Dimensions are in mm.*

The apparatus in Figure 15 is bolted to a steel base to support the assembly in a tank of water as shown in Figure 16. There is a thermocouple in the water tank to measure the temperature increase during a timed experiment. The water tank and the aluminum components are insulated. The mass of the aluminum base is 150 g and the steel base is 350 g.

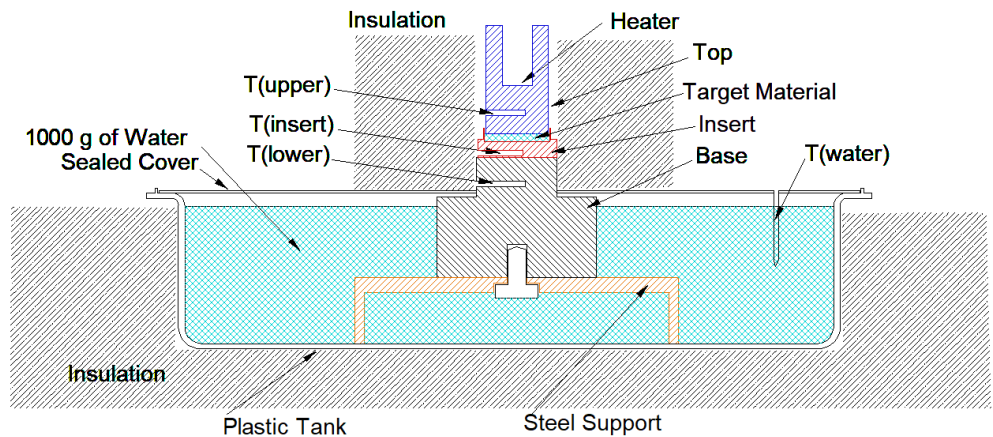

*Figure 16. Details of the setup used to measure the heat transfer through the target material in the insert.*

The materials tested were indium powder of 325 mesh, and $CaCO_3$ powder ground from chalk to about 200 mesh and several mixtures of composite targets of indium powders and $CaCO_3$. Tests were conducted with volume ratios of 50/50, 40/60 and 33/67 of indium and $CaCO_3$. Note that the volume ratio refers to the compacted material without voids and not to the loose powder. The results shown are for 40 % indium and 60 % $CaCO_3$. A four-sided glass container of about 4 cm each side with rounded corners was used to manually tumble mix the components for a composite target. The uniformity was judged visually.

## A2. Measurement Details

In a typical measurement, the powder was poured into the insert and leveled at about 4.0 mm thick. The mass of the insert was checked at the beginning of each test and the mass of the insert plus powder was also measured before test. The top piece was used to compress the mixture by a small amount. The insert and top piece were installed in the base and an insulated keeper was bolted to the base to hold the assembly together and provide a clamping force estimated to provide a pressure of up to 28 MPa. The heater was installed, and upper insulation was added. The heater was turned on for 900 s and temperatures were measured at the four locations shown in Figure 15, $T_{upper}$, $T_{insert}$, $T_{lower}$ and $T_{water}$, every 30 s for 400 s and every 60 s for another 1200 s, by which time the assembly had cooled to nearly the starting conditions. The temperatures measured using commercial digital thermometers with a resolution of 1 $^o$C and an estimated accuracy of about 0.4 $^o$C. In what follows, $\Delta T$ without a subscript will be calculated as

$$\Delta T = T_{upper} - T_{insert.} \qquad (5)$$

The water was stirred on an intermittent basis and the heater was turned off at 900 s. The energy, $Q$, transferred to the water can be obtained using

$$Q = \Delta T_w c_w m_w, \qquad (6)$$

where $\Delta T_w$ is the increase of the water temperature, $c_w$ is the specific heat capacity of water and $m_w$ is the mass of the water. There is an additional, smaller amount of energy, transferred to the aluminum and steel components during each measurement. This number divided by 900 s gives the average power transferred through the target material once an equilibrium temperature is reached. The water was at surrounding air temperature at the beginning of the test and typically increased about 6 $^o$C during a test. It was still within 0.2 $^o$C of that temperature at another 1000 s after the end of the test, indicating that the insulation was adequate.

The power produced by the heater was measured by measuring the temperature rise of 1000 cm$^3$ of water, heated for 900 s. The heater was installed directly in a hole in the aluminum top piece (see Figure 15) to minimize heat loss to the surroundings. The water temperature was measured to 0.1 $^o$C accuracy, each 60 s for 1500 s, by which time there was no further change. The mass of the aluminum base (150 g) and the steel support (350 g) heat by the same amount as the water by the end of the 1500 s. This test indicated that the water heated by 6.4 $^o$C, equivalent to 29.8 W during the 900 s time period and the metal components account for an additional 2.2 W or a total of 32.0 W. During most tests the power measured was close to this number.

The targets were examined after each test to check for things such as uniformity of mixed components, if it was compacted into a solid disk and adhesion to the insert and top piece. The top piece was removed from the insert plus target material and the mass was checked to determine if any material was lost between the start and end of the test. The height of the compressed target material was measured, and this was used with the area of the insert (about 2.8 cm$^2$) to determine the volume of the material as tested. The mass and volume of the material after the test was used to determine an approximate density of the composite target material. The mass measurement was accurate to 0.05 g with masses from typically 1.5 to 4.5 g, and the thickness measurement was accurate to about 0.05 mm from typical measurements from about 1.0 to 3.0 mm, leading to density measurements with an accuracy of 7 to 10 %.

The density of the two powders and one mixture were measured by weighing several samples of 4.7 cm$^3$ of each material. The densities with no compaction were: indium - 3.4 g/cm$^3$ or 47 % of solid density; CaCO$_3$ - 1.0 g/cm$^3$ or 38 % of solid density; 40/60 mixture of indium and CaCO$_3$ - 2.0 g/cm$^3$ or 44 % of solid density.

## A3. Results

### A3.1. Measurements with a 40/60 In /CaCO$_3$ Composite Target

A mixture of indium and CaCO$_3$ powders was tumble-mixed in a 4-sided glass container to reasonable consistency. The amount of each powder was chosen so that, when compacted, the ratio of indium to CaCO$_3$ would be 40 % to 60 % by volume. Figure 17 shows two measurements for different thicknesses. The thermal conductivity, $k$, can be obtained using

$$P = kA\Delta T_d/\Delta x, \tag{7}$$

where $P$ is the heat power passing through the sample, $A$ is the sample area and $\Delta T_d$ is the temperature drop across a sample of thickness $\Delta x$. There is a $\Delta T_d$ of about 11.5 $^o$C between the two samples and a $\Delta x$ of 14 x $10^{-4}$ m, leading to an estimate of the thermal conductivity of about 13.9 W/m-K.

In addition to the thermal conductivity of the target material, the thermal conductance, $h$, has an important impact on the temperature of the target material when irradiated. The pressure of the cooling water on the aluminum holder will be in the range of 1 to 2 MPa and the data in Figure 17 shows that $\Delta T$ in this case is about 31.0 $^o$C. $\Delta T$ can be written as

$$\Delta T = \Delta T_{Al} + \Delta T_{tgt} + 2\Delta T_{inf}, \tag{8}$$

Where $\Delta T_{Al}$, $\Delta T_{tgt}$ and $\Delta T_{inf}$ are the temperature drops across the aluminum, the target material and each interface, respectively. Using the known thermal conductivities and the heat flux of 11.4 W/cm$^2$, $\Delta T_{inf}$ can be calculated to be 3.3 $^o$C. Then $h$ is found to be about 3.5 W/cm$^2$-K. Note that, if the heat flux during irradiation is about 200 W/cm$^2$, the temperature step at the interface will be almost 60 $^o$C.

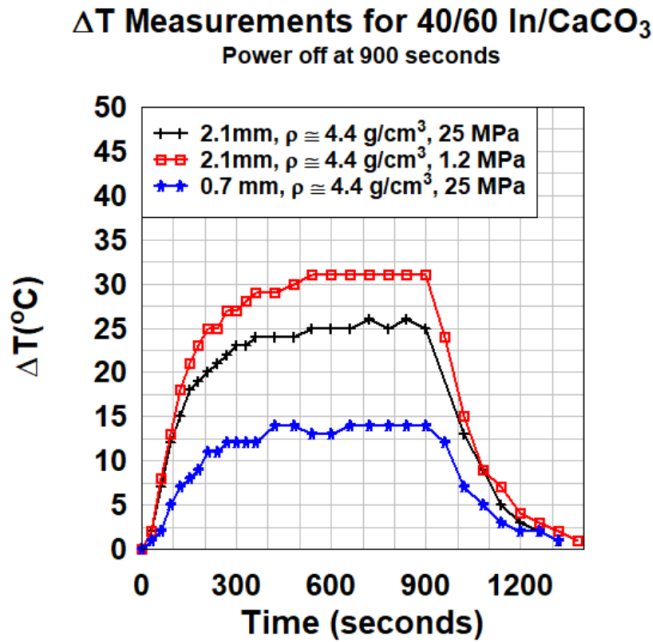

*Figure 17. Measured values of $\Delta T$ for composite targets of 40% indium and 60% CaCO$_3$ by volume.*

### A3.2. Measurements with Compacted Powdered CaCO$_3$

Figure 18 shows two measurements of $\Delta T$ for compacted CaCO$_3$ targets of different thicknesses (0.5 mm and 1.4 mm), loaded at approximately the same pressure of 18 MPa. Using Figure 2, this compaction pressure will lead to a relative density of about 0.83 or a porosity, $\varphi_v$, of 0.17.

For the 1.4 mm target, $\Delta T$ is about 139 $^o$C and for the 0.5 mm target, about 89 $^o$C. The temperature difference, $\Delta T_d$, between the two samples is about 50 $^o$C for a thickness difference of 0.9 mm and power of about 30 W or 10 W/cm$^2$. Using Equation (7) leads to a value of $k$ of 1.8 W/m-K, about 80 % of the value of the solid material. Using Equation (3) with a porosity of 0.17 indicates that the thermal conductivity would be about 70 % of the solid material, somewhat lower than measured. However, we should note that we have not measured the thermal conductivity of the solid material and there is considerable variation in values reported in the literature.

The blue curve in Figure 18 shows $\Delta T$ for the thin target at reduced loading. Using the same approach as for the In/CaCO$_3$ target, the thermal conductance can be estimated to be 0.22 W/cm$^2$-K. Note that this value is more than an order of magnitude smaller than for the In/CaCO$_3$ target.

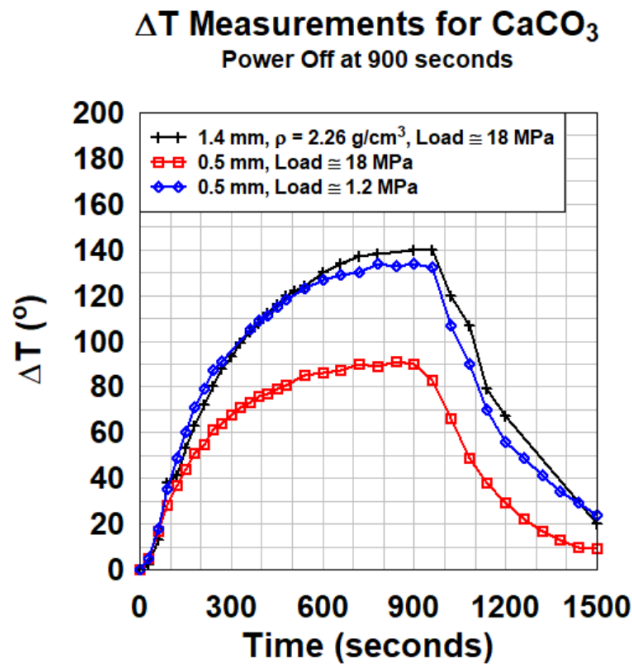

*Figure 18. Values of $\Delta T$ for tests of compacted CaCO$_3$ powder of different thicknesses.*

### A3.3. Measurements with Compacted Indium Powder

Figure 19 shows values of $\Delta T$ for two tests of compacted indium powder with different loading. The heater was only operated for 600 s (compared to 900 s for most tests) because $\Delta T$ reached equilibrium in about 200 s. At the end of the test the density was measured to be nearly equal to the metal value of 7.3 g/cm$^3$ and both metal pieces were adhered to the indium and required a significant force to separate them.

In this case, we do not have data with two different sample thicknesses in order to estimate the thermal conductivity using the temperature differential. Because of the limited resolution of our temperature probes and the small temperature drop across the pure indium sample we cannot extract a meaningful estimate of the thermal conductivity.

However, using the measured data at reduced loading and the accepted value of the thermal conductivity of pure indium we can use the same approach as for the In/CaCO$_3$ target to estimate the thermal conductance at the indium/aluminum interface. The result is 23 W/cm$^2$-K, about seven times larger than for the In/CaCO$_3$ sample. Because of our limited temperature resolution, this result has a significant uncertainty but is consistent with the use of indium to improve interface thermal conductance[30].

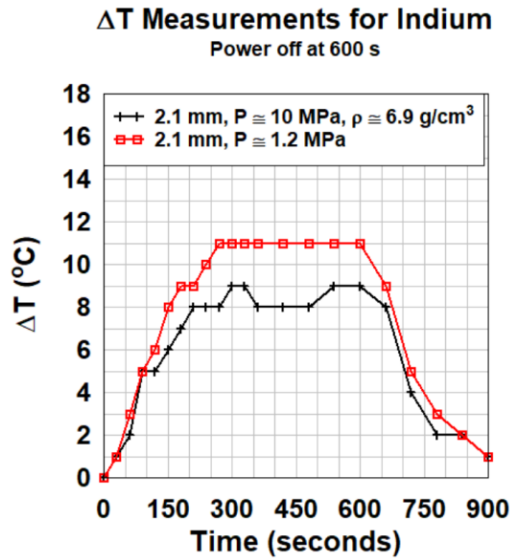

*Figure 19. Measured values of ΔT for indium powder at two different compaction pressures.*

## A4. Summary

A series of tests have been done with indium and CaCO$_3$ powders and composite targets with a ratio of 40 % to 60 % by volume of indium to CaCO$_3$. It was shown that target material with a relative density close to unity can be produced with modest pressures. The measured thermal conductivity is about 17 % of that of pure indium, with an estimated uncertainty of 10 % to 15 %. The thermal resistance presented by the interfaces between the target material and the aluminum holder will have a significant impact on the equilibrium temperature of the target material when irradiated with a high power beam.

## References

[1]   A. Jang, A. T. Kendi, G. B. Johnson, T. R. Halfdanarson, and O. Sartor, "Targeted Alpha-Particle Therapy: A Review of Current Trials," *Int. J. Mol. Sci.*, vol. 24, no. 14, p. 11626, Jul. 2023, doi: 10.3390/ijms241411626.

[2]   NorthStar Medical Radioisotopes, "Accelerating Production of Non-Carrier Added Actinium-225 (n.c.a. Ac-225)," White Paper, 2023. [Online]. Available: https://www.northstarnm.com/company/library/#whitepapers

[3]   W.-J. Lin and G. Harbottle, "Gamma ray emission intensities of $^{226}$Ra in equilibrium with its daughter products," *J. Radioanal. Nucl. Chem.*, vol. 153, no. 2, pp. 137–149, Feb. 1991, doi: 10.1007/BF02164874.

[4]   IAEA, "Cyclotron produced radionuclides: Principles and practice," IAEA, Vienna, Technical Report Series 465, 2008.

[5]   L. C. Wei, L. E. Ehrlich, M. J. Powell-Palm, C. Montgomery, J. Beuth, and J. A. Malen, "Thermal conductivity of metal powders for powder bed additive manufacturing," *Addit. Manuf.*, vol. 21, pp. 201–208, May 2018, doi: 10.1016/j.addma.2018.02.002.

[6]   X. Qu, L. Zhang, M. Wu, and S. Ren, "Review of metal matrix composites with high thermal conductivity for thermal management applications," *Prog. Nat. Sci. Mater. Int.*, vol. 21, no. 3, pp. 189–197, Jun. 2011, doi: 10.1016/S1002-0071(12)60029-X.

[7]   A. Stolarz *et al.*, "Targets for production of the medical radioisotopes with alpha and proton or deuteron beams," *AIP Conf. Proc.*, vol. 1962, no. 1, p. 020004, May 2018, doi: 10.1063/1.5035517.

[8]   Biobase, "BK-VX series - Orbital mixer by Biobase | MedicalExpo." https://www.medicalexpo.com/prod/biobase/product-84845-1124102.html (accessed Nov. 17, 2024).

[9]   H. J. Ryu, C. K. Kim, M. Sim, J. M. Park, and J. H. Lee, "Development of high-density U/Al dispersion plates for Mo-99 production using atomized uranium powder," *Nucl. Eng. Technol.*, vol. 45, no. 7, pp. 979–986, Dec. 2013, doi: 10.5516/NET.07.2013.014.

[10]  W. T. Diamond and C. K. Ross, "Actinium-225 production with an electron accelerator," *J. Appl. Phys.*, vol. 129, no. 10, p. 104901, Mar. 2021, doi: 10.1063/5.0043509.

[11]  B. Dionne, A. Bergeron, J. R. Licht, Y. S. Kim, and G. L. Hofman, "Thermal Properties for the Thermal-Hydraulics Analyses of the BR2 Maximum Nominal Heat Flux," Argonne National Lab. (ANL), Argonne, IL (United States), ANL/RERTR/TM-11-20REV.1, Feb. 2015. doi: 10.2172/1168941.

[12]  Agency for Toxic Substances and Disease Registry, *Toxicological profile for radium*. Altanta, GA: U.S. Department of Health and Human Services, 1990.

[13] P. J. Denny, "Compaction equations: a comparison of the Heckel and Kawakita equations," *Powder Technol.*, vol. 127, no. 2, pp. 162–172, Oct. 2002, doi: 10.1016/S0032-5910(02)00111-0.

[14] K. Pietrak and T. Wiśniewski, "A review of models for effective thermal conductivity of composite materials," *J. Power Technol.*, vol. 95, no. 1, pp. 14–24, 2014.

[15] R. Landauer, "The Electrical Resistance of Binary Metallic Mixtures," *J. Appl. Phys.*, vol. 23, no. 7, pp. 779–784, Jul. 1952, doi: 10.1063/1.1702301.

[16] J. K. Carson, S. J. Lovatt, D. J. Tanner, and A. C. Cleland, "Thermal conductivity bounds for isotropic, porous materials," *Int. J. Heat Mass Transf.*, vol. 48, no. 11, pp. 2150–2158, May 2005, doi: 10.1016/j.ijheatmasstransfer.2004.12.032.

[17] M. E. Cunningham and K. L. Peddicord, "Heat conduction in spheres packed in an infinite regular cubical array," *Int. J. Heat Mass Transf.*, vol. 24, no. 7, pp. 1081–1088, Jul. 1981, doi: 10.1016/0017-9310(81)90157-5.

[18] "Materials Database - Thermal Properties - Thermtest Inc.," *Thermtest*. https://thermtest.com/thermal-resources/materials-database (accessed Aug. 06, 2024).

[19] A. Ferrari, P. R. Sala, A. Fasso, and J. Ranft, "FLUKA: A multi-particle transport code (Program version 2005)," Oct. 2005, doi: 10.2172/877507.

[20] T. T. Böhlen *et al.*, "The FLUKA Code: Developments and Challenges for High Energy and Medical Applications," *Nucl. Data Sheets*, vol. 120, pp. 211–214, Jun. 2014, doi: 10.1016/j.nds.2014.07.049.

[21] C. Ahdida *et al.*, "New Capabilities of the FLUKA Multi-Purpose Code," *Front. Phys.*, vol. 9, 2022, Accessed: May 25, 2022. [Online]. Available: https://www.frontiersin.org/article/10.3389/fphy.2021.788253

[22] C. Apostolidis, R. Molinet, J. McGinley, K. Abbas, J. Möllenbeck, and A. Morgenstern, "Cyclotron production of Ac-225 for targeted alpha therapy," *Appl. Radiat. Isot.*, vol. 62, no. 3, pp. 383–387, Mar. 2005, doi: 10.1016/j.apradiso.2004.06.013.

[23] D. Horn *et al.*, "Actinium isotope cross sections for 226Ra(p,xn) reactions measured at low energies," *Appl. Radiat. Isot.*, vol. 212, p. 111427, Oct. 2024, doi: 10.1016/j.apradiso.2024.111427.

[24] G. F. Steyn *et al.*, "A vertical-beam target station and high-power targetry for the cyclotron production of radionuclides with medium energy protons," *Nucl. Instrum. Methods Phys. Res. Sect. Accel. Spectrometers Detect. Assoc. Equip.*, vol. 727, pp. 131–144, Nov. 2013, doi: 10.1016/j.nima.2013.06.041.

[25] G. Steyn, C. Vermeulen, and E. Isaacs, "Encapsulation methods for solid radionuclide production targets at a medium-energy cyclotron facility," *AIP Conf. Proc.*, vol. 1962, no. 1, p. 030016, May 2018, doi: 10.1063/1.5035533.

[26] A. K. H. Robertson, C. F. Ramogida, P. Schaffer, and V. Radchenko, "Development of $^{225}$Ac Radiopharmaceuticals: TRIUMF Perspectives and Experiences," *Curr. Radiopharm.*, vol. 11, no. 3, p. 156, Dec. 2018, doi: 10.2174/1874471011666180416161908.

[27] Elmer, "Elmer multiphysical simulation software." [Online]. Available: https://research.csc.fi/web/elmer/elmer

[28] C. V. Madhusudana, *Thermal contact conductance*, Second. Switzerland: Springer International, 2014.

[29] Dassault Systémes, "Thermal Contact Resistance - 2022 - SOLIDWORKS Help." https://help.solidworks.com/2022/english/SolidWorks/cworks/c_Thermal_Contact_Resistance.htm (accessed Nov. 27, 2024).

[30] B. Jarrett and J. Ross, "Comparison of test methods for high-performance thermal interface materials," Indium Corporation, Technical Paper.

[31] J. B. Ainscough, "Gap conductance in Zircaloy-clad LWR fuel rods," OECD Nuclear Energy Agency, Paris, CSNI Report 72, 1982.

[32] Z. Ge, D. G. Cahill, and P. V. Braun, "Thermal Conductance of Hydrophilic and Hydrophobic Interfaces," *Phys. Rev. Lett.*, vol. 96, no. 18, p. 186101, May 2006, doi: 10.1103/PhysRevLett.96.186101.

[33] H. S. Huang, A. K. Roy, V. Varshney, J. L. Wohlwend, and S. A. Putnam, "Temperature dependence of thermal conductance between aluminum and water," *Int. J. Therm. Sci.*, vol. 59, pp. 17–20, Sep. 2012, doi: 10.1016/j.ijthermalsci.2012.04.016.

[34] W. R. Gambill and R. D. Bundy, "Heat-Transfer Studies of Water Flow in Thin Rectangular Channels: Part I - Heat Transfer, Burnout, and Friction for Water in Turbulent Forced Convection," *Nucl. Sci. Eng.*, vol. 18, no. 1, pp. 69–79, Jan. 1964, doi: 10.13182/NSE64-A18141.

[35] Grove Software, "Radiation Software." https://radiationsoftware.com/microshield (accessed May 24, 2024).

[36] Sino Santech, "Indium - Sino SanTech." https://www.sinosantech.com/indium (accessed May 24, 2024).

[37] TMAX, "Lab 12T Manual Hydraulic Press." https://www.tmaxlaboratory.com/lab-12t-manual-hydraulic-press-with-a-digital-pressure-gauge-optional-commonly-used-in-infrared-laboratories_p90.html (accessed Nov. 17, 2024).

[38] Mettler-Toledo, "METTLER TOLEDO Balances & Scales for Industry, Lab, Retail." https://www.mt.com/ca/en/home.html (accessed Nov. 17, 2024).

[39] Gabrian International, "6061 Aluminum Alloy - Properties".

[40] *Shigley's Mechanical Engineering Design*. 2019. Accessed: Nov. 07, 2024. [Online].
Available: https://www.mheducation.com/highered/product/shigley-s-mechanical-
engineering-design-budynas-nisbett/M9780073398211.html